\documentclass[aps,prl,reprint,superscriptaddress,longbibliography,nofootinbib]{revtex4-1}

\usepackage{amsfonts}
\usepackage{amsmath}
\usepackage{tikz}
\usetikzlibrary{arrows,calc}
\usetikzlibrary{patterns,snakes}

\DeclareMathOperator{\Tr}{Tr}

\newcommand{\bea}{\begin{eqnarray}}
\newcommand{\eea}{\end{eqnarray}}
\newcommand{\Op}{{\mathcal O}}

\begin{document}

%\preprint{}

\title{Generalized Eigenstate Thermalization in 2d  CFTs}


\author{Anatoly Dymarsky}
\affiliation{Department of Physics and Astronomy, \\ University of Kentucky, Lexington, KY 40506\\[2pt]}
\affiliation{Skolkovo Institute of Science and Technology, \\ Skolkovo Innovation Center, Moscow, Russia, 143026\\[2pt]}
\author{Kirill Pavlenko}
\affiliation{Skolkovo Institute of Science and Technology, \\ Skolkovo Innovation Center, Moscow, Russia, 143026\\[2pt]}
\affiliation{Moscow Institute of Physics and Technology, \\ Dolgoprudny, Russia, 141700 \\[2pt]}


\date{\today}
\begin{abstract}
Infinite-dimensional conformal symmetry in two dimensions leads to  integrability of 2d conformal field theories  by giving rise to an infinite tower of local conserved qKdV charges in involution. We discuss how presence of  conserved charges constraints equilibration in 2d CFTs. We propose that in the thermodynamic limit large central charge 2d CFTs satisfy generalized eigenstate thermalization, with the values of qKdV charges forming a complete set of thermodynamically relevant quantities, which unambiguously determine expectation values of all local observables from the vacuum family.   Equivalence of ensembles further provides that local properties  of an eigenstate can be described by the Generalized Gibbs Ensemble that only includes qKdV charges. In the case of a general initial state, upon equilibration, emerging Generalized Gibbs Ensemble will necessary include negative chemical potentials  and holographically will be described by a quasi-classical black hole with quantum soft hair. 
\end{abstract}

%Our work settles the question if non-local or quasi-local charges are necessary to describe equilibrium of large central charge 2d CFTs.

%slow thermalization modes

%\keywords{}

\maketitle

The topic of thermalization, and more generally, equilibration of isolated many-body quantum systems has been an active area of research during the past decade. In case of non-integrable systems, i.e.~those without an extensive number of local conserved quantities, emergence of the thermal equilibrium has been traced to eigenstate thermalization hypothesis (ETH) which postulates thermal properties of individual energy eigenstates \cite{deutsch1991quantum,srednicki1994chaos,rigol2008thermalization}. In the simplest form it requires the expectation value of some appropriate (often taken to be local) observable $\Op$ in a many-body eigenstate $|E_i\rangle$ to be a smooth function of energy,
\begin{equation}
\label{eth}
\langle E_i|\Op|E_i\rangle=f_\Op(E_i).
\end{equation}
Qualitatively, eq.\eqref{eth}  postulates that energy is the only thermodynamically relevant quantity, which completely specifies local properties of an eigenstate. 
The condition \eqref{eth} may apply to all or most eigenstates, in which case it is referred as strong or weak ETH. The eigenstate thermalization ensures equivalence between the expectation value in the eigen-ensemble, $f_\Op(E_i)$, and thermal expectation value of $\Op$ in the Gibbs ensemble, $f_\Op(E_i)=\Tr(e^{-\beta H}\Op)/Z$, where the effective temperature $\beta$ is fixed through the energy balance relation, $E_i=\Tr(e^{-\beta H}\Op)/Z$ \cite{d2016quantum}.

When the system is integrable, with an extensive number of conserved charges  $Q_i$, ETH does not apply. Accordingly emerging equilibrium can be different from the Gibbs state. In this case  the equilibrium can be described by the Generalized Gibbs Ensemble (GGE), a generalization of grand canonical ensemble that includes an infinite tower of conserved charges \cite{rigol2007relaxation}. Validity of the GGE has been related to the generalized eigenstate thermalization \cite{cassidy2011generalized,he2013single,vidmar2016generalized}, which generalizes \eqref{eth} to include an infinite number of conserved quantities,
\bea
\label{geth}
\langle E_i|\Op|E_i\rangle=f_\Op(Q_k(E_i)).
\eea 
Here $|E_i\rangle$ is a mutual eigenstate of the Hamiltonian and charges $Q_k$, $Q_k(E_i)$ are the eigenvalues of $Q_k$ associated with $|E_i\rangle$, and function $f_\Op$ is assumed to be a smooth function of  all of its arguments. Similarly to \eqref{eth}, at the qualitative level, \eqref{geth} postulates that charges $Q_k$ form a complete set of thermodynamically relevant quantities which fully specify local properties of an eigenstate. Provided \eqref{geth} applies to most states, it ensures equivalence between the generalized microcanonical ensemble and GGE, establishing validity for the latter to describe emerging equilibrium e.g.~following a quantum quench  \cite{cassidy2011generalized}.

In this Letter we discuss thermalization of two-dimensional conformal field theories (CFTs), a rich topic with multiple connections ranging from the cold atom experiments \cite{calabrese2016quantum} to physics of quantum gravity \cite{anous2016black}. It has been shown that following a quantum quench 2d conformal theories equilibrate and reach a steady state, which in many cases can be described in terms of the Gibbs ensemble \cite{calabrese2006time,calabrese2007quantum}. At the same time emergence of thermal equilibrium is not universal.
Conformal symmetry in two dimensions gives rise to an infinite tower of local mutually commuting  conserved qKdV charges $Q_{2k-1}$, the CFT Hamiltonian for the left-movers being $Q_1\equiv H$, a part of the integrable structure of the 2d CFTs \cite{bazhanov1996integrable,bazhanov1997integrable,bazhanov1999integrable}. The question we are concerned with  is how presence of these charges affects equilibration. By analogy with the integrable lattice models it is natural to expect that locally equilibrium states can be described in terms of the GGE, which includes all local qKdV charges. Indeed, emergence of exactly such qKdV GGE was analytically shown  for a special family of so-called Cardy-Calabrese initial states \cite{cardy2016quantum}. %An important question emphasized by their work is to establish or rule out the need to include other charges besides the local qKdV ones to GGE to describe equilibrium  in $c\gg 1$ CFTs. 

In the context of integrable systems  the question which quantities should be included in the GGE is far from being trivial. Early studies in the context of XXZ and Lieb-Liniger models  have shown that a full set of extensive local charges does not specify local properties of eigenstates, signaling failure of generalized ETH \cite{brockmann2014quench,mestyan2015quenching}. These works raised an important question of the validity of the GGE to describe an emerging equilibrium following a quantum quench \cite{wouters2014quenching,pozsgay2014correlations,goldstein2014failure}. A resolution comes from the fact that besides local conserved quantities these models give rise to quasi-local conserved charges \cite{ilievski2015quasilocal}.  Taking them into account restores validity of the GGE \cite{ilievski2015complete}. %Quench action 
Following studies in the context of integrable field theoretic models, both free and interacting ones, have decisively established that adding quasi-local charges is necessary to accurately describe the after-quench equilibrium state \cite{essler2015generalized,sotiriadis2016memory,doyon2017thermalization,bastianello2017quasi,palmai2018quasilocal}.  These findings raise an important question emphasized in \cite{cardy2016quantum} if the set of local qKdV charges is generally sufficient to describe equilibrium in large $c$ 2d CFTs, or it should be extended by non-local or perhaps some new local charges \cite{vernier2017quasilocal}. In this Letter we answer this question by showing that at large $c$, 2d CFTs satisfy generalized eigenstate thermalization \eqref{geth} with the local qKdV charges forming a complete set and ambiguously specifying local properties of the eigenstates.

Two-dimensional conformal field theories admit a split into non-interacting sectors of left and right movers. For simplicity we only discuss one sector explicitly, while all results automatically extend to the full theory.  We consider 2d CFT on a circle of the circumference $\ell$ in a mutual eigenstate of all charges $Q_{2k-1}$,
\bea
\label{state}
|E\rangle= |\{m_i\},\Delta\rangle,\quad E=Q_1=(\Delta+\sum m_i)/\ell,
\eea
labeled by the primary state $\Delta$ and the set of integers $\{m_i\}$ \cite{GGE}. The set $\{m_i\}$ is convenient to parametrize using free boson representation where an integer $n_k$ for $k=1,2,\dots$ specifies the number of times integer $k$ appears in the set $\{m_i\}$. In the thermodynamic limit $\ell\rightarrow \infty$  ``energy'' $Q_1$ and all other qKdV charges are assumed to scale with the system size to yield finite charge densities \
%$q_1=E/\ell$,
$q_{2r-1}=Q_{2r-1}/\ell$. In terms of $\Delta, n_k$ this implies scaling 
\bea
\Delta \sim \ell^2,\quad \sum_k n_k\, k^{2r-1} \sim \ell^{2r}.
\eea 
In what follows we restrict the discussion to the eigenstates \eqref{state} with the density  charges 
$q_{2r-1}=\langle E|Q_{2r-1}|E\rangle/\ell$ which additionally satisfy 
\bea
\label{qcondition}
 {q_{2r-1}\over q_1^r}=1+O(1/c).
\eea
Here and below the CFT central charge $c$ is assumed to be large. 
Holographically, this regime corresponds to a quasi-classical black hole in $AdS_3$, where one in the RHS of \eqref{qcondition} corresponds to classical gravity, while $O(1/c)$ term is due to quantum corrections \cite{de2016remarks,perez2016boundary,GGE,maloney2018generalized}. In terms of $\Delta,n_k$ an exponential majority of states in the generalized microcanonical ensemble specified by $q_{2k-1}$  subject to \eqref{qcondition}  will satisfy
\bea
\label{restriction}
{\sum_k n_k\, k^{2r-1} \over \Delta^r}=O(1/c^{r}).
\eea
In fact \eqref{restriction} may apply to all states in the generalized  microcanonical ensemble \eqref{qcondition}, yielding  strong version of the generalized ETH in 2d CFTs.  
To verify that one would need to know full spectrum of qKdV charges, going beyond currently known leading $1/c$ expansion.

In the regime of quasi-classical gravity \eqref{restriction}, $c\gg 1$, expectation values of qKdV charges can be calculated explicitly \cite{GGE2},
\bea
\label{weak1}
\ell q_1&=&\Delta +\sum_k n_k k,\\
\label{weak2}
\ell q_3&=&\Delta^2 +\sum_k n_k \left(6\Delta k+ {c\, k^3 \over 6}\right)+O(c^0),\\
\dots \nonumber \\
\label{weak3}
\ell q_{2r-1}&=&\Delta^r+\sum_k n_k\, p_{2r-1}(c,\Delta, k)+O(c^{r-2}),
\eea
where $p_{2r-1}(c,\Delta, k)$ are some known polynomials of degree $2r-1$ which include only odd powers of $k$. 

Because of translational invariance the expectation value of a full derivative $\Op=\partial \mathcal O'$ in energy  eigenstate will vanish. Hence it suffices to consider  expectation values $\langle E|{\mathcal O}|E\rangle$  only when ${\mathcal O}$ is a quasi-primary operator. Below we consider the case when ${\mathcal O}$ belongs to the vacuum family, i.e.~it is a Virasoro descendant of the identity. 
%The case when ${\mathcal O}$ is a descendants of other primary operators is discussed at the end of this letter. Because of translational invariance expectation value of $\Op$ is equal to the integral 
%\bea
%\langle {\mathcal O}\rangle={1\over \ell} \int_0^\ell du\, \langle E|{\mathcal O}(u)|E\rangle,
%\eea
%which often simplifies calculations. 
To streamline the notations we  introduce $\langle {\mathcal O}\rangle\equiv \langle E|{\mathcal O}|E\rangle$.
It is convenient to parametrize $\Op$ by its dimension (level). At the levels $2$ and $4$ there are unique quasi-primaries in the vacuum family, 
\bea
\Op_2=T,\qquad \Op_4=T^2-{3\over 10}\partial^2 T.
\eea
Thus expectation values of $\Op_{2,4}$ are identically equal to charge densities $q_1,q_3$ \cite{lashkari2018universality}.
At the level $6$ there are two quasi-primaries (we always choose quasi-primaries in the basis which diagonalizes Zamolodchikov metric)
\bea
\Op_6^{(1)}&=&T^3 - \frac{9}{10} (T \partial^2 T) + \frac{4}{35} \partial^4 T + \frac{93}{70c + 29}\Op_6^{(2)},\ \ \\
\Op_6^{(2)}&=&(\partial T \partial T) - \frac{4}{5}(T \partial^2 T) + \frac{23}  {210} \partial^4 T.
\eea
The expectation value of the combination $\Op_6^{(1)}+\frac{5}{9} \frac{c}{12}\Op_6^{(2)}$ is identically  equal to $q_5$. Similarly to (\ref{weak1}-\ref{weak3}), at leading order the expectation value of $\Op_6^{(2)}$ has the form of a polynomial in $\Delta$ and odd powers of $k$,
\bea
\langle\Op_6^{(2)}\rangle=\frac{9}{5}\sum_k n_k \left(\frac{c}{6} k^5  + 4\Delta k^3\right)+O(c^0).
\eea
It is possible to use (\ref{weak1}-\ref{weak3}) to express any term of the form  $\sum_k n_k\, k^{2r-1}$ via $q_{2j-1}$, $j\leq r$, but a priori the result would also depend on $\Delta$. Thus, at leading order in $1/c$, expectation values of $\Op_6^{(i)}$ are some functions of $\Delta$ and $q_{2r-1}$. Remarkably, because of the non-trivial cancellations the final result is $\Delta$-independent, 
and can be expressed solely in terms of $q_{2r-1}$. To simplify the answer we introduce dimensionless ratio ${\rm q}_{2k-1}=q_{2k-1}/q_1^k$ such that $\delta {\rm q}_{2k-1}\equiv {\rm q}_{2k-1}-1$ is of order $1/c$. Then $\Op_6^{(i)}$ measured in units of energy density $q_1$ is given by 
%\bea
%\langle\Op_6^{(2)}\rangle= \frac{9}{5}\frac{12}{c} \left(q_5 - q_1^3 - 3 q_1 (q_3 - q_1^2) \right)+O(c^0).\ \ \
%\eea
%\bea
%q_{1}^{-3}\langle\Op_6^{(2)}\rangle=\frac{108}{5c} \left(\delta {\rm q}_5 - 3\, \delta {\rm q}_3 \right)+O(1/c^{3}).\ \ \
%\eea
\bea
q_1^{-3}\langle {\mathcal O}_6^{(1)} \rangle &=& 1 + 3\, \delta {\rm q}_3 +O(1/c^2), \\
q_1^{-3}\langle {\mathcal O}_6^{(2)} \rangle &=& \frac{9}{5}\frac{12}{c} \left(\delta{ \rm q}_5 - 3\, \delta{\rm q}_3  \right)+O(1/c^3).
\eea
As we see different quasi-primaries have different scaling with $c$. Our calculation applies to leading $1/c$ behavior of each quasi-primary, except for a special one, which includes maximal power of  $T$ without derivatives. The expectation value of that quasi-primary starts with $O(c^0)$ and our result applies to the first two terms in $1/c$ expansion.

The possibility to express eigenstate expectation value $\langle \Op\rangle$ as a polynomial in $q_{2j-1}$ extends to all higher levels. 
For an operator of dimension $2r$ the answer only depends on $q_{2j-1}$ for $j\leq r$.
We write down explicit expressions for all operators up to level $10$ in terms of $q_{2j-1}$ in Supplemental Materials. Our results establish generalized eigenstate thermalization for vacuum block observables in large $c$ CFTs.

That expectation value $\langle \Op\rangle$ of an operator of dimension $2r$ only includes qKdV charges $q_{2j-1}$ up to the same dimension $j\leq r$ can be interpreted as a manifestation of locality. It is analogous 
to the observation in the context of integrable lattice models that to describe equilibrium state locally, at the length scales not exceeding some distance $a$, 
it is only necessary to include local and quasi-local charges in the GGE with the support within $a$ \cite{nandy2016eigenstate,pozsgay2017generalized}.

%physics upon equilibration of some initial state $|\Psi\rangle$, 

Generalized eigenstate thermalization implies  validity of the qKdV Generalized Gibbs Ensemble 
\bea
\label{gge}
\rho={\rm exp}\left\{-\sum_k \mu_{2k-1} Q_{2k-1}\right\}/Z,\quad \mu_1\equiv\beta,
\eea 
to describe local properties of individual energy eigenstates, 
provided chemical potentials $\mu_{2k-1}$ are tuned to match values of the eigenstate charges 
\bea
\label{Legendre} 
\ell \,q_{2k-1}=\langle E_i|Q_{2k-1}|E_i\rangle=\Tr(\rho\, Q_{2k-1}).
\eea
Provided $q_{2k-1}$ a chosen to represent charge densities of some non-equilibrium initial state $|\Psi\rangle$, 
a standard argument would consequently equate the GGE expectation values of local operators with those in the diagonal ensemble of $|\Psi\rangle$, written in the eigenbasis \eqref{state}. In most cases the latter would be equal to the
expectation values in state $|\Psi\rangle$ upon equilibration. It should be noted though that left and right Hamiltonians $Q_1,\bar{Q}_1$ are highly degenerate, and therefore validity of the diagonal ensemble to describe local physics upon equilibration may be violated.  

It remains an open question to establish existence of $\mu_{2k-1}$ which would solve \eqref{Legendre} for any given set of $q_{2k-1}$. Using explicit form of the generalized partition function in the large $c$ limit \cite{GGE2} we can find, up to $O(1/c^2)$ corrections, 
\bea
\label{disc}
&&\delta{\rm q}_{2k-1}={q_{2k-1}\over q_1^k}-1=\\ &&{24k\over c}\int_0^\infty {d\kappa \kappa \left[(2k-1){}_2F_1(1,1-k,3/2,-\kappa^2)-1\right] \over e^{2\pi \kappa \gamma}-1}, \nonumber \\
&& \gamma=\sum_{j=1}^\infty \tilde{\mu}_{2j-1}j(2j-1){\sigma}^{j-1/2}{}_2F_1(1,1-j,3/2,-\kappa^2)\nonumber,
\eea
where $\tilde{\mu}_{2k-1}={\sqrt{6}\over \pi}c^{k-1}\mu_{2k-1}$ and $\sigma(\tilde{\mu}_{2k-1})$ is positive and satisfies
\bea
\sum_{k=1} k\,\tilde{\mu}_{2k-1}\, \tilde{\sigma}^{k-1/2}=1.
\eea
From here it follows that when all chemical potentials are positive $q_{2k-1}$ satisfy an infinite series of inequalities (see Supplemental Materials)
\bea
\label{inequalities}
&&{q_{3}\over q_1^2}-1\leq {22\over 5c}+O(1/c^2),\quad  {q_{5}\over q_1^3}-1\leq {302\over 21c}+O(1/c^2),\nonumber \\   && \dots 
\eea
Thus GGE emerging after equilibration of some general initial state will have to include negative chemical potentials, unless all inequalities \eqref{inequalities} are satisfied. 

\begin{figure}[t]
\includegraphics[width=0.5\textwidth]{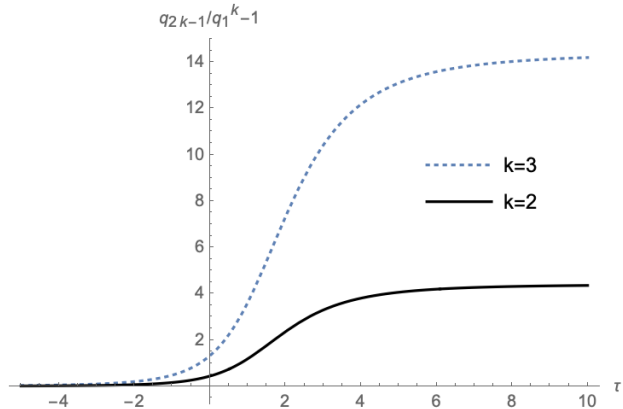}
\caption{Plot of $q_{2k-1}/q_1^k-1$ in the units of $1/c$ as a function of $\tau=\beta(\pi^2/(6c\mu_3))^{1/3}$ for $k=2,3$. It approaches zero as $|\tau|^{-3}$ for all $k$ when $\tau\rightarrow -\infty$. The opposite limit $\tau\rightarrow \infty$ corresponds to the Gibbs ensemble, $q_1\sim \beta^{-1}$, $\mu_3\rightarrow 0$, and $c(q_{2k-1}/q_1^k-1)$ for $k=2,3$ approach $22/5$ and $302/21$ correspondingly.} 
\label{Figure}
\end{figure}

To match GGE to a primary state all qKdV densities should be related to each other via $q_{2k-1}=q_1^k$ \cite{lashkari2018universality}. This is only possible if the integral in  
 \eqref{disc} vanishes, which requires $\gamma$ to be infinite. This is consistent with the observation of \cite{GGE} that an ensemble with any finite number of non-zero $\mu_{2k-1}$ can not describe primary states. This is because in full generality $q_{2k-1}\geq q_1^k$ and hence primary states are at the boundary of the phase space of $q_{2k-1}$'s. 
It is nevertheless possible to describe them in the limit, via a GGE with at lest some coefficients approaching infinity. The simplest scenario is to consider $\mu_3>0$ and arbitrary $\beta\equiv \mu_1$, while all other chemical potentials are identically zero. Then in the limit $\tau = \beta(6/\pi^2c\mu_3)^{1/3}\rightarrow -\infty$, for all $k$, $q_{2k-1}/q_1^k-1$ will vanishes as $\sim |\tau|^{-3}$, as is shown  for $k=2,3$ in Fig.~\ref{Figure}.

With just two chemical potentials $\beta,\mu_3$ being non-zero the values of $q_{2k-1}/q_1^k-1$ is confined to be between zero and their thermal (Gibbs ensemble) values. This constraint is removed already after turning on  one more additional chemical potential. For example by taking $\beta,\mu_5>0$ and $\mu_3<0$ one can fine-tune function $\gamma$ to become arbitrarily small for some positive value of $\kappa$, leading to the divergence of the integral in \eqref{disc} and violating quasi-classical regime \eqref{qcondition}.

From the holographic point of view equilibration in field theory is associated with the formation of a black hole in $AdS_3$, a background dual to the GGE \eqref{gge}.   Conserved qKdV charges correspond to the black hole soft hair, which are only visible at quantum level. At the level of classical gravity $c\rightarrow \infty$ all qKdV charges are related, $q_{2k-1}=q_1^k$. Accordingly there is a unique classical BTZ black hole family of solutions parametrized by $q_1,\bar q_1$ \cite{de2016remarks,perez2016boundary}. It is an important question to understand the regime $q_{2k-1}\neq q_1^k$ holographically, by including quantum gravity corrections into consideration. This, in particular, should provide holographic interpretation to negative temperature and other chemical potentials, which will necessarily appear starting from a general initial state. 

In this Letter we have only considered local probes $\mathcal O$ from the vacuum block. In case $\mathcal O$ is a non-trivial Virasoro primary, or its descendant, it will have zero expectation value in the GGE \eqref{gge} for any values of $\mu_{2k-1}$. This is because in the thermodynamic limit $\ell\rightarrow \infty$ geometry degenerates into a cylinder, which is conformally flat. Thus, to satisfy any version of eigenstate thermalization the eigenstate expectation value $\langle E|{\mathcal O}|E\rangle$ must simply vanish. In terms of the CFT data, this means most or all heavy-heavy-light Operator Product Expansion coefficients must approach zero when the dimension of heavy operators grows to infinity. If that is the case, generalized eigenstate thermalization will be trivially  satisfied. It remains an outstanding problem to establish if large central charge chaotic CFT, in particular those with gravity duals,  exhibit this behavior. 

{\it Conclusions.} In this Letter we have established that large central charge 2d CFTs  in the thermodynamic limit satisfy generalized eigenstate thermalization with the tower of local qKdV charges forming a complete set of thermodynamically-relevant quantities. Our analysis establishes universal validity of Generalized Gibbs Ensemble that includes all qKdV charges to describe
individual energy eigenstates, and hence in most cases, asymptotic equilibrium states in such theories.  It would be important to extend the analysis to next order in $1/c$, which will likely reveal if the eigenstate thermalization is strong i.e.~applies to all finite energy density eigenstates, or weak, i.e.~applies to most states. %Another important question is to understand the role of qKdV charges from the holographic point of view.\\
 
\acknowledgments
We thank Alex Avdoshkin, Dmitry Abanin, Tomaz Prosen, Marcos Rigol, and Alexander Zhiboedov for discussions. 

\bibliography{GETH}

\newpage
\section{Supplemental Materials}
\subsection{Expectation value of quasi-primaries in eigenstates}
In this section we list the explicit expressions for the eigenstate expectation values of all quasi-primaries up to level ten in terms of qKdV charges. 
\subsubsection{Level 6}
There is are two quasi-primaries 
\bea
\Op_6^{(1)}&=&T^3 - \frac{9}{10} (T \partial^2 T) + \frac{4}{35} \partial^4 T + \frac{93}{70c + 29}\Op_6^{(2)},\ \ \ \\
\Op_6^{(2)}&=&(\partial T \partial T) - \frac{4}{5}(T \partial^2 T) + \frac{23}  {210} \partial^4 T.
\eea
In the limit (\ref{restriction}) they can be simplified to
\bea
\Op_6^{(1)}&=&T^3 + O(1/c), \\
\Op_6^{(2)}&=&\frac{9}{5}(\partial T \partial T) + O(1/c).
\eea
In units of the energy density their expectation values are
\bea
q_1^{-3}\langle {\mathcal O}_6^{(1)} \rangle &=& 1 + 3\, \delta {\rm q}_3 +O(1/c^2), \\
q_1^{-3}\langle {\mathcal O}_6^{(2)} \rangle &=& \frac{9}{5}\frac{12}{c} \left(\delta{ \rm q}_5 - 3\, \delta{\rm q}_3  \right)+O(1/c^3).
\eea

\subsubsection{Level 8}
There are three quasi-primaries at level 8, 
\bea
\Op_8^{(1)}&=&T^4 + O(1/c), \\
\Op_8^{(2)}&=&\frac{9}{5}(T(\partial T \partial T)) + O(1/c), \\
\Op_8^{(3)}&=& \frac{143}{63}(\partial^2 T \partial^2 T) + O(1/c).
\eea
In  the units of energy density at leading order they are
\bea
q_1^{-4}\langle \Op_8^{(1)} \rangle &=& 1 + 6 \, \delta{\rm q}_3+O(1/c^2), \\
q_1^{-4}\langle \Op_8^{(2)} \rangle &=& \frac{9}{5} \frac{12}{c} \left(  \delta{\rm q}_5 - 3\,\delta{\rm q}_3  \right)+O(1/c^3),\\
q_1^{-4}\langle \Op_8^{(3)} \rangle &=& \frac{143}{63} \frac{180}{c^2} \left( \delta{\rm q}_7 - 4\, \delta{\rm q}_5  + 6\,\delta{\rm q}_3   \right)+O(1/c^4). \nonumber
\eea
\subsubsection{Level 9}
There are no quasi-primaries of odd dimension smaller than nine. At level nine there is a unique quasi-primary ${\mathcal O}_9$,
%\bea
%{\mathcal O}_9=
%\eea
which has zero expectation value, as well as all higher odd-dimensional quasi-primaries, due to parity. 

\subsubsection{Level 10}
There are four quasi-primaries at level 8. In the  limit  (\ref{restriction})  up to some additional factors they are 
\bea
\Op_{10}^{(1)}&=&T^5 + O(1/c), \\
\Op_{10}^{(2)}&=&(T(T(\partial T \partial T))) + O(1/c),\\
\Op_{10}^{(3)}&=&(T(\partial^2 T \partial^2 T)) + O(1/c), \\
\Op_{10}^{(4)}&=& (\partial^3 T \partial^3 T) + O(1/c).
\eea
In terms of energy density 
their expectation values are
\bea
q_1^{-5}\langle \Op_{10}^{(1)} \rangle &=& 1 + 10\, \delta{\rm q}_3+ O(1/c^2), \\
q_1^{-5}\langle \Op_{10}^{(2)} \rangle &=& \frac{1}{c} \left(  \delta{\rm q}_5 - 3\,\delta{\rm q}_3  \right)+ O(1/c^3), \\
q_1^{-5}\langle \Op_{10}^{(3)} \rangle &=& \frac{180}{c^2} \left(  \delta{\rm q}_7 - 4\, \delta{\rm q}_5 + 6\, \delta{\rm q}_3  \right)+ O(1/c^4), \\
q_1^{-5}\langle \Op_{10}^{(4)} \rangle &=& \frac{3024}{c^3} \left(  \delta{\rm q}_9 - 5\, \delta{\rm q}_7 + 10\,\delta{\rm q}_5 - 10\, \delta{\rm q}_3   \right)+ O(1/c^5).\nonumber
\eea

\subsection{GGE with positive chemical potentials}
For any positive integer $j$ hypergeometric function ${}_2F_1(1,1-j,3/2,-\kappa^2)$ is polynomial in $\kappa^2$ with non-negative coefficients which starts with one, 
\bea
{}_2F_1(1,1-j,3/2,-\kappa^2)=1+{2\over 3}(j-1)\kappa^2+\dots
\eea
Hence it  is a monotonically increasing function of $\kappa$ which satisfies ${}_2F_1(1,1-j,3/2,-\kappa^2)\geq 1$. From here it follows that when all chemical potentials are non-negative, function  $\gamma$ defined in the equation \eqref{disc} from the main text satisfies 
\bea
\nonumber
\gamma\geq \sum_{j=1}^\infty \tilde{\mu}_{2j-1}j(2k-1)\sigma^{j-1/2}\geq \sum_{j=1}^\infty \tilde{\mu}_{2j-1}j \sigma^{j-1/2}=1.
\eea
Thus at leading order in $1/c$, $q_{2k-1}/q_1^k-1$ is bounded from above by its value in the Gibbs ensemble,
\bea
&&\delta {\rm q}_{2k-1}\leq \\ \nonumber
&&{24k\over c}\int_0^\infty {d\kappa \kappa \left[(2k-1){}_2F_1(1,1-k,3/2,-\kappa^2)-1\right] \over e^{2\pi \kappa}-1}=\\
&& {k\over c}\left(\sum_{p=0}^{k-1} {6(2k-1) \Gamma(k)\Gamma(1/2)\over \Gamma(p+3/2)\Gamma(k-p)} (-1)^{p+1}\zeta(-1-2p)-1\right). \nonumber
\eea
This yields $22/5$ for $k=2$, $302/11$ for $k=3$, $2428/75$ for $k=4$, and so on. \\

\subsection{GGE with two non-zero chemical potentials}
To gain better intuition it is instructive to consider the generalized ensemble which includes only two charges, the conventional Hamiltonian of CFT $H\equiv Q_1$ and $Q_3$,
\bea
\rho={\rm exp} \left(-\beta H-\mu_3 Q_3\right)/Z.
\eea
To assure convergence we must require $\mu_3>0$ while $\beta$ can be arbitrary. It is convenient to parametrize $\beta,\mu_3$ in terms of 
\bea
\tau=\beta \left({6\over \pi^2c \mu_3}\right)^{1/3}, 
\eea
and energy density $q_1=-\ell^{-1}{\partial \ln Z\over \partial \beta}$, such that
\bea
\nonumber
\beta&=&q_1^{-1/2}\left({c\pi^2\over 6}\right)^{1/2}\frac{\tau  \left(\sqrt[3]{\tau ^3+3 \left(\sqrt{6 \tau ^3+81}+9\right)}-\tau \right)}{\sqrt{6} \sqrt[6]{\tau ^3+3 \left(\sqrt{6 \tau ^3+81}+9\right)}},\\
\mu_3&=&q_1^{-3/2}\left({c \pi^2\over 6}\right)^{1/2}\frac{\left(\sqrt[3]{\tau ^3+3 \left(\sqrt{6 \tau ^3+81}+9\right)}-\tau \right)^3}{6 \sqrt{6} \sqrt{\tau ^3+3 \left(\sqrt{6 \tau ^3+81}+9\right)}}. \nonumber
\eea
Then $\delta {\rm q}_{2k-1}$ only depends on $\tau$,
\bea
\nonumber
&&\gamma=1+\frac{2^{1/2} \left(\kappa ^2+1\right) \left(\sqrt[3]{\tau ^3+3 \left(\sqrt{6 \tau ^3+81}+9\right)}-\tau \right)^3}{3^{3/2} \sqrt{\tau ^3+3 \left(\sqrt{6 \tau ^3+81}+9\right)}},\\
&&\delta {\rm q}_{2k-1}=\\[4pt] \nonumber
&&\qquad {24 k\over c}\int_0^\infty {d\kappa\,\kappa\left[(2k-1){}_2F_1(1,1-k,3/2,-\kappa^2)-1\right]\over e^{2\pi \kappa \gamma}-1}.
\eea
When $\tau$ approaches minus infinity while $q_1$ is kept fixed,
\bea
&& \beta\sim - q_1^{-1/2}\left({c \pi^2\over 6}\right)^{1/2}|\tau|^{3/2}2^{-1/2},\\
&& \mu_3\sim  q_1^{-3/2}\left({c \pi^2\over 6}\right)^{1/2}|\tau|^{3/2}2^{-3/2},
\eea
and we find that $c \delta {\rm q}_{2k-1}$ approaches zero as $1/|\tau|^3$.
We plot $\delta {\rm q}_{2k-1}\equiv q_1^{-k}q_{2k-1}-1$ in the units of $1/c$ as a function of $\tau$ for $k=2,3$ in Fig.~\ref{Figure} in the main text. 

\end{document}